\documentclass[apjl, iop]{emulateapj}

\newcommand  \mgii  {\ifmmode {\rm Mg}{\textsc{ii}} \else Mg\,{\sc ii}\fi}
\newcommand  \MGII  {\ifmmode {\rm Mg}\,{\sc ii}\,\lambda2798 \else Mg\,{\sc ii}\,$\lambda2798$\fi}
\newcommand  \siiv  {\ifmmode {\rm Si}\, {\sc iv}\ \else Si\,{\sc iv}\fi}
\newcommand  \SIIV  {\ifmmode {\rm Si}\,{\sc iv}\,\lambda1399 \else Si\,{\sc iv}\,$\lambda1399$\fi}
\newcommand  \cv  {\ifmmode {\rm C}\, {\sc v}\ \else C\,{\sc v}\fi}
\newcommand  \aliii  {\ifmmode {\rm Al}{\textsc{iii}} \else Al\,{\sc iii}\fi}
\newcommand  \civ  {\ifmmode {\rm C}\, {\sc iv}\ \else C\,{\sc iv}\fi}
\newcommand  \ciii  {\ifmmode {\rm C}\, {\sc iii}\ \else C\,{\sc iii}\fi}
\newcommand  \oiii  {\ifmmode \left[{\rm O}\,{\textsc iii}\right] \else [O\,{\sc iii}]\fi}
\newcommand  \CIV  {\ifmmode {\rm C}\,{\sc iv}\,\lambda1549 \else C\,{\sc iv}\,$\lambda1549$\fi}
\newcommand  \NV  {\ifmmode {\rm N}\,{\sc v}\,\lambda1240 \else N\,{\sc v}\,$\lambda1240$\fi}
\newcommand  \nv  {\ifmmode {\rm N}\,{\sc v}\ \else N\,{\sc v}\fi}
\newcommand  \ovi    {\ifmmode \left[{\rm O}\,{\textsc vi}\right] \else O\,{\sc vi}\fi}
\newcommand  \LyA  {\ifmmode {\rm Lyman}\,{\sc $\alpha$}\,\lambda1216 \else Lyman\,{\sc $\alpha$}\,$\lambda1216$\fi}
\newcommand  \lya {\ifmmode {\rm Lyman}\,{\sc $\alpha$}\ \else Lyman\,{\sc $\alpha$}\fi}
\newcommand  \ha {\ifmmode {\rm H~}\,{\sc $\alpha$}\ \else H\,{\sc $\alpha$}\fi}
\newcommand  \feii     {Fe\,{\sc ii}}

\newcommand  \ALIII  {\ifmmode {\rm Al}\,{\sc iii}\,\lambda1857 \else Al\,{\sc iii}\,$\lambda1857$\fi}

\newcommand{\kms}{\ifmmode {\rm km\,s}^{-1} \else km\,s$^{-1}$ \fi}

\def\sulii{[S\,{\sc ii}]}
\def\nii{[N\,{\sc ii}]}

\shorttitle{ BAL wind shedding dust cocoon }
\shortauthors{Yi et al.}

\begin{document}


\title{ A quasar shedding its dust cocoon at redshift 2 } 

\author{ Weimin Yi\altaffilmark{1,2,12}, W.~N. Brandt\altaffilmark{2,3,4}, Q. Ni\altaffilmark{2}, Luis C. Ho\altaffilmark{5,6}, Bin Luo\altaffilmark{7}, Wei Yan\altaffilmark{2}, D.~P. Schneider\altaffilmark{2,3}, Jeremiah D. Paul\altaffilmark{8}, Richard M. Plotkin\altaffilmark{8}, Jinyi Yang\altaffilmark{9}, Feige Wang\altaffilmark{9}, Zhicheng He\altaffilmark{10}, Chen Chen\altaffilmark{11}, Xue-Bing Wu\altaffilmark{5,6}, Jin-Ming Bai\altaffilmark{1,12} } 

\altaffiltext{1}{Yunnan Observatories, Chinese Academy of Sciences, Kunming, 650216, China; wmyi2012@gmail.com }
\altaffiltext{2}{Department of Astronomy \& Astrophysics, The Pennsylvania State University, 525 Davey Lab, University Park, PA 16802, USA}  
\altaffiltext{3}{Institute for Gravitation and the Cosmos, The Pennsylvania State University, University Park, PA 16802, USA}
\altaffiltext{4}{Department of Physics, 104 Davey Laboratory, The Pennsylvania State University, University Park, PA 16802, USA}
\altaffiltext{5}{Kavli Institute for Astronomy and Astrophysics, Peking University, Beijing 100871, China}
\altaffiltext{6}{Department of Astronomy, Peking University, Yi He Yuan Lu 5, Hai Dian District, Beijing 100871, China}
\altaffiltext{7}{School of Astronomy and Space Science, Nanjing University, Nanjing, Jiangsu 210093,  China}
\altaffiltext{8}{Department of Physics, University of Nevada, Reno, NV 89557, USA} 
\altaffiltext{9}{Steward Observatory, University of Arizona, Tucson, AZ, 85721-0065, USA}
\altaffiltext{10}{CAS Key Laboratory for Research in Galaxies and Cosmology, Department of Astronomy, University of Science and Technology of China, Hefei, Anhui 230026, China}
\altaffiltext{11}{School of Physics \& Astronomy Sun Yat-Sen University, Zhuhai 519000,  China} 
\altaffiltext{12}{Key Laboratory for the Structure and Evolution of Celestial Objects, Chinese Academy of Sciences, Kunming 650216, China}

\begin{abstract}
We present the first near-IR spectroscopy and joint analyses of multi-wavelength observations for SDSS J082747.14+425241.1, a dust-reddened, weak broad emission-line quasar (WLQ) undergoing a remarkable broad absorption line (BAL) transformation. The systemic redshift is more precisely measured to be $z=2.070\pm0.001$ using H$\beta$ compared to $z=2.040\pm0.003$ using \mgii\ from the literature, signifying an extreme \mgii\ blueshift of $2140\pm530$ \kms\ relative to H$\beta$. Using the H$\beta$-based single-epoch scaling relation with a  systematic uncertainty of 0.3 dex, its black hole (BH) mass and Eddington ratio are estimated to be $M_{\rm BH}\sim6.1\times10^8M_\odot$ and $\lambda_{\rm Edd}\sim0.71$, indicative of being in a rapidly accreting phase. Our investigations confirm the WLQ nature and the LoBAL$\rightarrow$HiBAL transformation, along with a factor of 2 increase in the \mgii+\feii\ emission strength and a decrease of 0.1 in $E(B-V)$ over two decades. The kinetic power of this LoBAL wind at $R\sim$15 pc from its BH is estimated to be $\sim$43\% of the Eddington luminosity, sufficient for quasar feedback upon its host galaxy albeit with an order-of-magnitude uncertainty. This quasar  provides a clear example of the long-sought scenario where LoBAL quasars are surrounded by dust cocoons, and wide-angle nuclear winds play a key role in the transition for red quasars evolving into the commonly seen blue quasars. 
\end{abstract}


\section{Introduction} 
\label{intro_sec}

Broad emission line (BEL) blueshifts relative to systemic redshift are often seen in quasars and  interpreted as evidence for quasar winds (e.g., \citealp{Gaskell82,Richards11,Yi20,Zuo20,Xu20}). As a comparison, broad absorption lines (BALs; \citealp{Weymann91}) are unambiguous evidence for outflows driven by quasars (quasar winds in short), presumably because of their extreme outflow properties that cannot be explained by any stellar processes known. There are two major BAL-type quasars, namely high-ionization BAL (HiBAL, typically traced by \nv, \civ, and/or \siiv) and low-ionization BAL (LoBAL, typically traced by \mgii\ and/or \aliii\ in addition to HiBAL species at the same velocity) quasars. The observed HiBAL/LoBAL fractions are $\sim$15\%/$\sim$1.5\% based on  optically identified quasar samples, but the intrinsic BAL fraction may be up to $\sim$40\% after selection corrections (e.g., \citealt{Allen11}).

Orientation and evolution are the two most widely accepted interpretations with respect to BAL phenomena. 
A growing number of  studies have suggested  that HiBAL and non-BAL quasars could represent different views of the same underlying quasar population, on the basis of both populations having many similarities in outflow and physical properties (e.g., \citealp{Rankine20,Xu20,Yi20,LiuB21}). This finding is more consistent with the orientation than evolution in accounting for the HiBAL phenomena. Conversely, BAL (particularly LoBAL) quasars have also been widely interpreted as being in a special evolutionary phase, in that high column-density BALs signify powerful winds predominantly driven by rapidly accreting SMBHs. Such powerful winds  support the long-sought blowout scenario for the origin of the observed large population of blue quasars. The two  interpretations of BAL phenomena may be reconciled when noticing that the orientation and evolution scenarios have been drawn mostly from HiBAL and LoBAL quasars, respectively.

There are evident differences between HiBAL and LoBAL populations. 
Previous observations have established that reddening decreases from LoBAL, HiBAL, to non-BAL quasars  (\citealp{Richards03}). While BAL quasars appear to be a minority of the entire quasar population, the BAL fraction increases as the increase of reddening (e.g., \citealp{Richards03}); in particular, some studies found that the observed BAL fraction is strikingly higher in red quasars than in blue quasars ($\gtrsim$50\% vs. $\sim$15\%; e.g., \citealp{Urrutia09, Fynbo13}). Interestingly, \citet{Hamann19} revealed from a large sample that LoBAL quasars have on average larger velocities, higher column densities, and softer ionizing spectral energy distributions (SEDs) than HiBAL quasars, suggesting the presence of extremely powerful outflows in the LoBAL population. In addition, \citet{Yi19a}  discovered a remarkable time-dependent trend of BAL variability in a LoBAL sample, such that weakening BALs outnumber strengthening BALs  on longer sampling timescales (see Fig.~9 in that work). As a comparison, this trend appears not to hold in the HiBAL population (e.g., \citealp{Filizak13, WangT15, DeCicco18, Rogerson18}).

The differences between the HiBAL and LoBAL populations above suggest an evolutionary path from red to blue quasars. Combined with BAL-variability studies from the literature, \citet{Yi21} found an overall LoBAL$\rightarrow$HiBAL/non-BAL transformation sequence along with a decrease in dust, shedding light on the evolutionary path from red to blue quasars,  i.e., a substantial fraction of LoBAL quasars could be caught in the act of casting off their dust cocoons (e.g., \citealp{Boroson92, Voit93,Glikman12}). Furthermore, this  scenario lends support to  the argument for dusty winds as the origin of quasar reddening (\citealt{Rivera21}).  However, some recent studies of luminous type-1 quasars found apparent diverging evidence:  there are no significant differences in  emission-line outflow properties between red and blue quasars matched in luminosity (e.g., \citealp{Temple19,Villar-Martin20,Fawcett22}).

Given the above statistical evidence for red quasars evolving into blue quasars and the lack of significant differences in emission-line outflow properties, it is meaningful to test the evolution path via individual objects, especially those being in a special phase and/or having unambiguous quasar winds. These extreme cases can be used to extend dynamic ranges of fundamental properties and relationships found from statistical studies. SDSS J082747.14+425241.1 (hereafter J0827) is one  such example characterized by remarkable BAL transitions occurring in \mgii, \aliii, and \civ\ BAL species, which offers a unique opportunity to investigate  BAL transformations. On the other hand, J0827 is a WLQ whose nature remains poorly understood.   
A puzzle of WLQs is that they appear to be normal quasars in nearly every aspect except for their unusually weak high-ionization BELs  and remarkable X-ray properties (e.g., \citealp{Fan99,Diamond09,WuJ12,Luo15,NiQ18,NiQ20,Timlin20}). The co-existence of weak BELs and transient BALs in J0827  provides a natural laboratory to explore the underlying link.

In this work, we present  new observational results and implications from joint analyses of optical/near-IR spectroscopy for this red quasar which has extreme properties reported in previous studies. Throughout this work, a flat cosmology with $H_0$ = 71 km s$^{-1}$Mpc$^{-1}$,  $\Omega_M$ = 0.27 and  $\Omega_{\Lambda}$ = 0.73 is adopted.  

\section{Previous and new observations}

\begin{table}
\centering
 \caption{  Spectroscopic observations of J0827}
 \begin{tabular}{lcccc}
  \hline\noalign{\smallskip}
Instrument & $\lambda$/$\Delta\lambda$ &  Spectral  &  Exp  &  Observation  \\
Name &  &  Coverage  &   &  Date \\
           &  &  ($\mu$m)  &  (hr)  &  (MJD) \\
  \hline\noalign{\smallskip}
(A)\\
SDSS & 1800 & 0.38--0.92 & 2.5 & 52,266 \\
SDSS & 1800 & 0.38--0.92 & 1.0 & 54,524 \\
BOSS & 1800 & 0.36--1.03 & 1.3 & 55,513 \\
BOSS & 1800 & 0.36--1.03 & 1.5 & 57,063 \\
HET/LRS2 & 2000 & 0.36--0.7 & 0.7 & 58,212  \\
HET/LRS2 & 2000 & 0.36--0.7 & 0.7 & 58,428  \\ 
Gemini/GMOS & 1200 & 0.37--1.03 & 2.5 & 58,486  \\
\hline
(B)\\

P200/TripleSpec & 2700 & 1.0--2.5 & 1.5 & 58,726  \\
Gemini/GNIRS & 800 & 1.0--2.5 & 2.0 & 58,920  \\
HET/LRS2 & 2000 & 0.36--0.7 & 0.7 & 59,140  \\
HET/LRS2 & 2000 & 0.36--0.7 & 0.7 & 59,498  \\
  \noalign{\smallskip}\hline
\end{tabular}
\label{table1}
\tablecomments{ (A) Previous observations of this quasar (see \citealp{Yi19b,Yi21}). (B) New observations of this quasar in this work.  }
\end{table}

As reported in \citet{Yi19b} and \citet{Yi21}, the four different-epoch spectra of this quasar from the Sloan Digital Sky Survey (SDSS; \citealt{York2000})  demonstrated  gradual BAL disappearance in multiple ions from MJD = 52,266 to 57,063; as a comparison, later observations from MJD = 58,212 to 58,486 by the Low-Resolution Spectrograph-2 (LRS2; \citealp{Chonis14}) mounted on the Hobby-Eberly Telescope (HET; \citealp{Ramsey98,Hill21}) and the GMOS mounted on the Gemini-North telescope clearly revealed the rapid emergence of a higher-velocity \civ\ BAL.

To trace the subsequent variability of this quasar, we obtained additional optical spectra using the HET/LRS-2. Unlike the discovery paper from \citet{Yi19b}, the new HET/LRS-2 spectra in this work were processed by the pipeline including flux calibration (\citealp{Davis18,Indahl19}),  with a typical uncertainty of $\sim$20\% in the absolute flux calibration.

In order to more precisely  determine its systemic redshift, we also performed near-IR spectroscopic observations of J0827 using 
the Palomar Hale 200 inch telescope (P200/TripleSpec; \citealt{Wilson04}). TripleSpec provides a wavelength coverage from 1.0 $\mu$m to 2.5~$\mu$m at an average spectral resolution of $\sim2700$, allowing simultaneous observations in the J/H/K bands. A slit width of one arcsecond and the ABBA dither pattern along the slit were chosen to improve the sky subtraction.

However, due to difficulties of analyzing the data with relatively low signal-to-noise ratio (S/N) obtained from P200/TripleSpec, we requested additional, higher-S/N near-IR  observations using the Gemini Near-IR Spectrograph (GNIRS) via the Fast Turnaround program at Gemini Observatory (PI: W. Yi). 
Observations of a standard star with coordinates close to the target were also executed to improve flux calibration and telluric correction. The Gemini spectra were processed using the Gemini IRAF package via standard techniques,  which have  an uncertainty range of 10--20\% in  flux calibration since they were taken in a non-photometric condition. The log of all the observations for this quasar analyzed in this study is shown in Table~\ref{table1}.

\section{Observational results} 
The remarkable optical properties of this quasar have been reported in \citet{Yi19b} and partly revisited in \citet{Yi21}, which will be mentioned only in passing in this work. Here, we mainly focus on delivering  new results based on the joint analyses of optical/near-IR spectroscopy.

\subsection{ Systemic redshift } \label{systemic_redshift}

Both near-IR spectra obtained by P200/TripleSpec and Gemini/GNIRS show strong, broad H$\alpha$ emission lines; furthermore, the higher-S/N Gemini/GNIRS spectrum  exhibited a narrow H$\beta$ component, producing a systemic redshift of $z=2.070\pm0.001$, a value that is in excellent agreement with that determined by the narrow H$\alpha$ emission peak. Throughout this work, we use only the Gemini/GNIRS spectrum for the related analyses and visualization. 

Noticeably, the H$\beta$-based systemic redshift is much higher than the \mgii-based systemic redshift $z\approx2.040\pm0.003$ that was determined from optical spectroscopy, implying a large \mgii\ blueshift (see Section~\ref{mgii_bel_var}) and higher line-of-sight (LOS) velocities than previously reported for the two BALs, i.e., the newly emerged \civ\ BAL falls into the range of extremely high-velocity (0.1--0.2$c$) outflows (\citealt{Paola20}).

\subsection{SED properties}
\label{SED_comparison}

\begin{figure} 
\center{}
 \includegraphics[height=7cm, width=8.6cm,  angle=0]{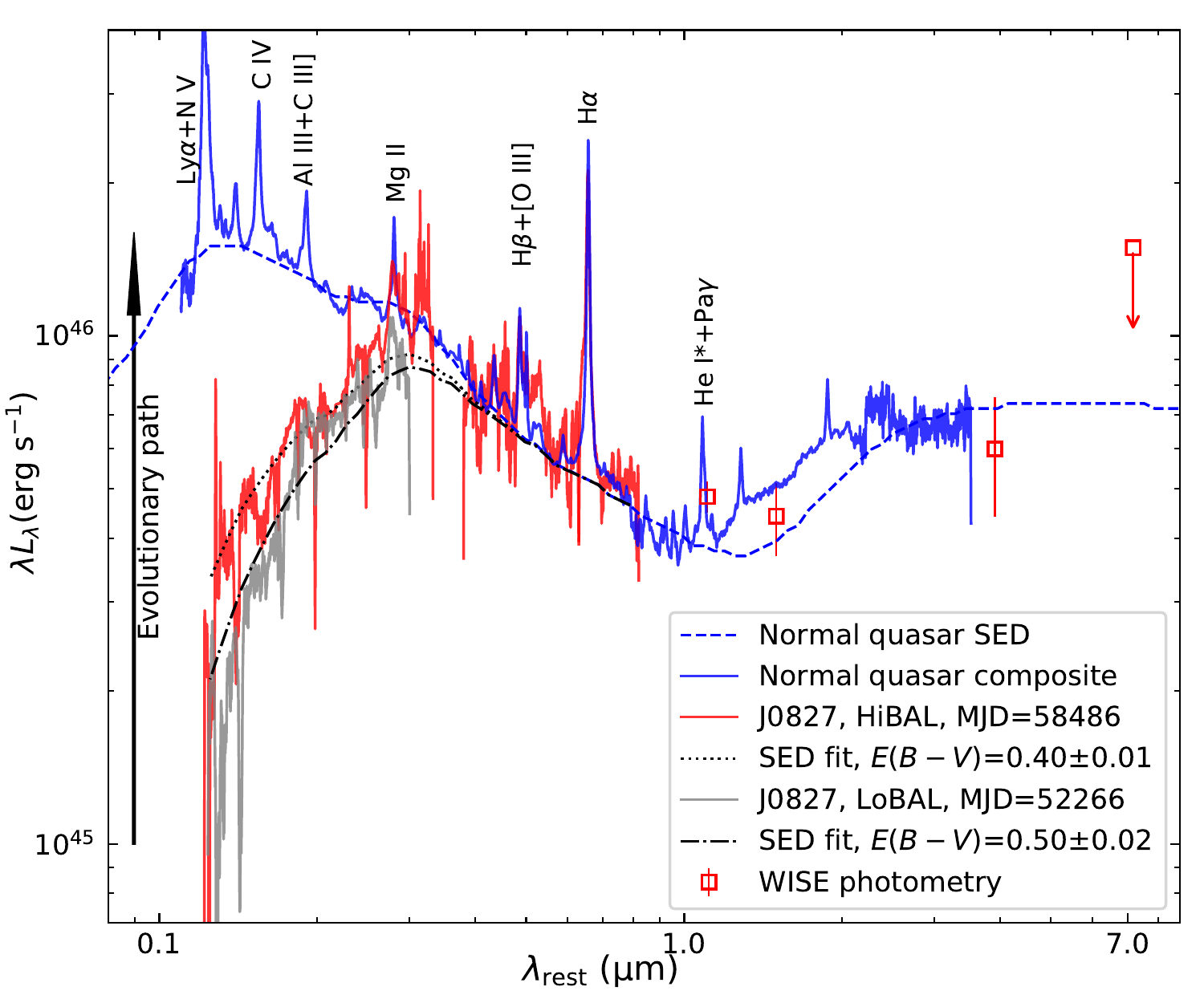} 
      \caption{   Comparisons between the LoBAL and non-BAL quasar composite spectra (scaled to the W1/W2/W3 fluxes of J0827). The squares are the photometric data from \textit{WISE}.  Here only the two-epoch spectra of J0827 are shown to depict the LoBAL$\rightarrow$HiBAL transformation along with a large decrease in reddening over the two epochs, consistent with the evolutionary path from red to blue quasars driven by quasar winds.  }
      \label{J0827sed_nonBAL2021Nov}
\end{figure}
To complete the optical/near-IR spectroscopic data, we searched for mid-IR photometric detections from the \textit{WISE} database (\citealt{Wright2010}) and found that J0827 was significantly detected ($>5\sigma$) in W1, W2, and W3, but only marginally  ($<3\sigma$) in W4.


Of particular interest here is to investigate the SEDs and emission-line properties between this (red) quasar and normal (blue) quasars (see Fig.~\ref{J0827sed_nonBAL2021Nov}). We construct the normal quasar composite spectrum from the literature (at $\lambda_{\rm rest}<0.3~\mu$m from \citealp{VandenBerk01} and at $\lambda_{\rm rest}>0.3~\mu$m from \citealp{Glikman06}). 
As shown in Fig.~\ref{J0827sed_nonBAL2021Nov}, there are several dramatic differences: (1) J0827 is heavily reddened compared to the non-BAL quasar composite at $\lambda_{\rm rest}<0.3~\mu$m, confirming a large amount of dust in J0827.  (2) The continuum of J0827 becomes  bluer at MJD=58486 than at MJD=52266, consistent with the evolutionary path from red to blue quasars. (3) There is no significant difference in H$\alpha$ emission  between J0827 and the normal quasar composite, in agreement with the finding from \citet{Banerji15}. (4) The SED of J0827 appears to peak at $\lambda_{\rm rest}=7~\mu$m, a  feature usually seen in red quasars. Although the W4-band detection for J0827 is marginal  and sets only an upper limit, the excess emission at $\lambda_{\rm rest}=7~\mu$m likely exists in J0827 (see Section~\ref{discussion_sec}). 

Assuming the scaled normal quasar SED (\citealt{Richards06}) in Fig.~\ref{J0827sed_nonBAL2021Nov} is the intrinsic SED of J0827, we perform the SED fits at $0.13<\lambda_{\rm rest}<0.8~\mu$m to quantify the change in continuum reddening over the two epochs. We choose the unusual reddening curve from \citet{JiangP13} rather than the widely adopted Small Magellan Cloud (SMC) reddening curve, as it can better reproduce the observed UV/optical continua (see Section~\ref{discussion_sec}). Specifically, we performed the spectral fit by modeling a reddened quasar SED to the relatively line-free windows identified by visual inspection, and the fitting uncertainties were generated by the standard deviations from the fits of 100 mock spectra via a Monte Carlo approach for both spectra.   

\subsection{Measurements from near-IR spectroscopy}
\label{BH_mass}

\begin{figure}
\center{}
 \includegraphics[height=6cm,width=8.6cm,  angle=0]{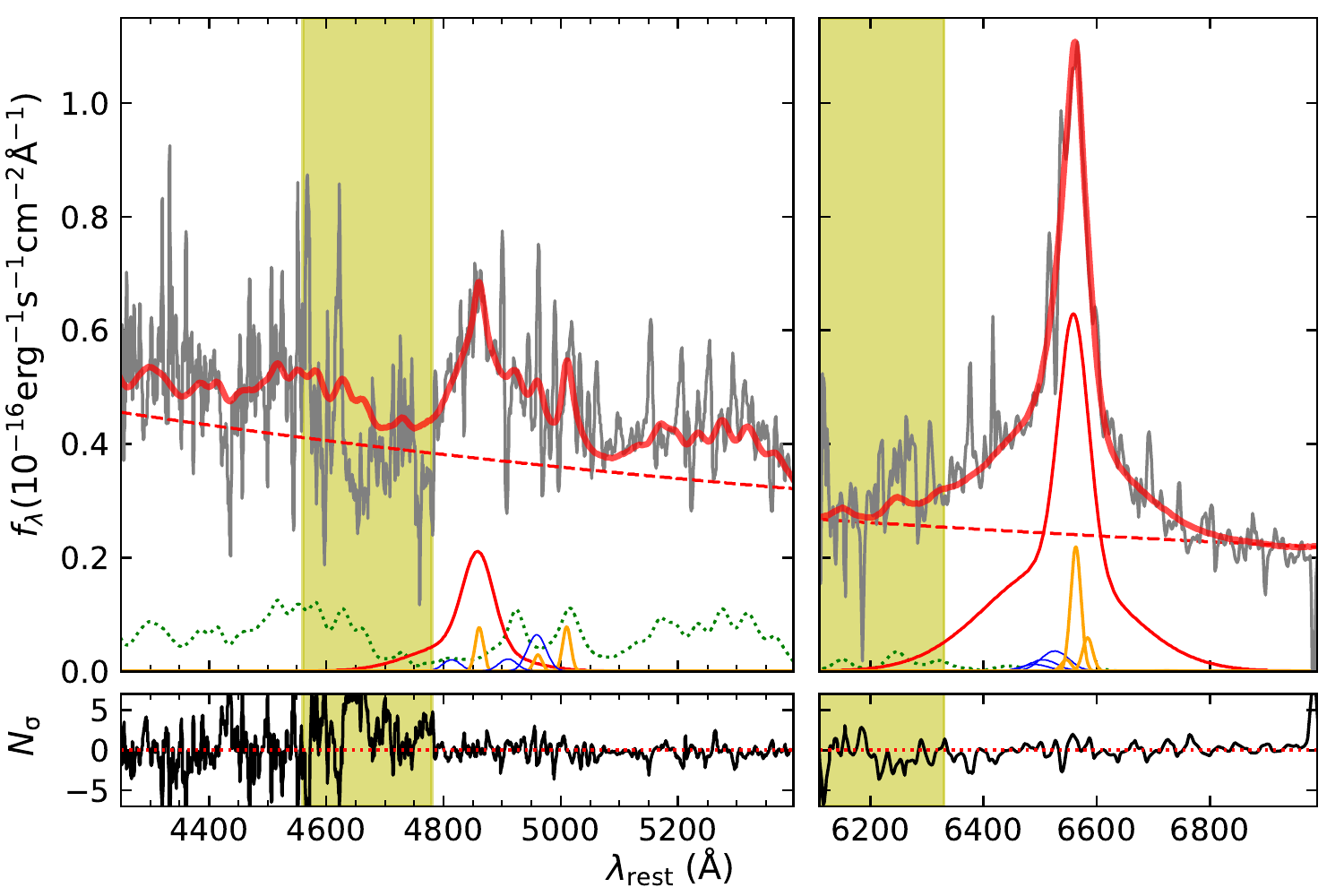} 
      \caption{  Spectral fits for the H$\beta$ (left) and H$\alpha$ (right) emission lines, in which the thick red lines are the models including the power-law continua (red dashed), the \feii\ components (green dotted), the narrow-core components (orange), the broad-wing components (blue), and the BLR components (thin red).  $N_{\sigma}$ in the bottom rows is the model-minus-data residual divided by the corresponding spectral error. Strong telluric absorption (yellow shaded) regions were masked during the fits. }
      \label{J0827HaHb_fitting}
\end{figure}

\begin{figure*}
\center{}
 \includegraphics[height=12cm,width=18cm,  angle=0]{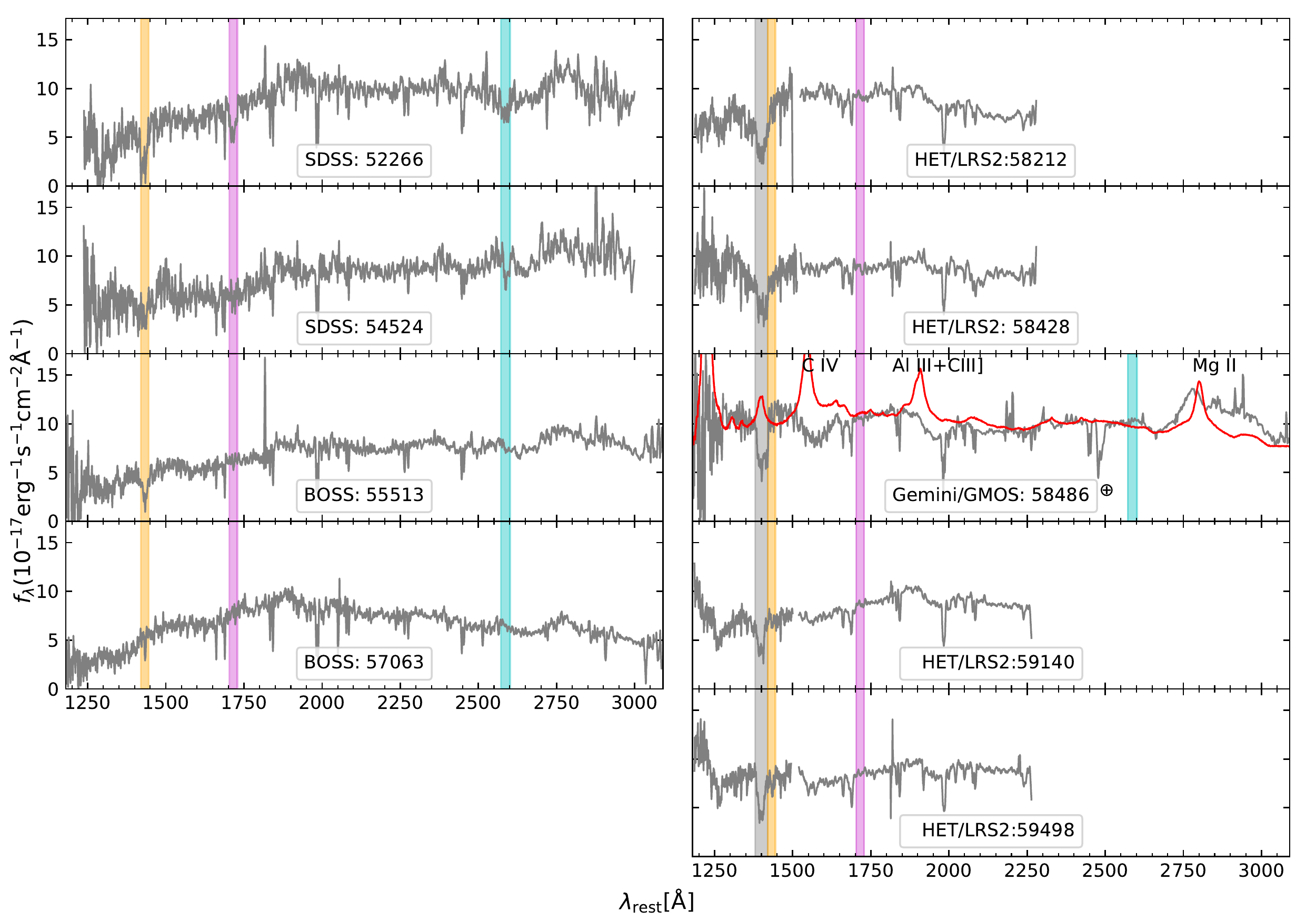} 
      \caption{ All the optical spectra (smoothed by a 15-pixel Savitzky-Golay filter) obtained by SDSS, Gemini/GMOS, and HET/LRS-2. The cyan, magenta, and orange shadings refer to the \mgii, \aliii, and \civ\ BALs at the same LOS-velocity range while the gray shading depicts the newly emerged \civ\ BAL at a higher LOS velocity. The red spectrum is the normal quasar composite from \citet{VandenBerk01} reddened by an unusual extinction curve from \citet{JiangP13} to match the spectrum at MJD = 58,486; the plus is for telluric absorption. It is clear that the \mgii/\aliii\ LoBALs (cyan/magenta shadings) completely disappeared at MJD = 57,063 and never reappeared in later epochs, while a HiBAL (gray shadings) emerged rapidly and is persistent after MJD = 57,063. During the LoBAL$\rightarrow$HiBAL transformation, the quasar continuum becomes progressively bluer in the later epoch while its \civ\ emission remains weak.  }
\label{J0827all_spec_fit_gmos4}
\end{figure*}

We choose H$\beta$ as the BH-mass tracer throughout this work, given that it has been confirmed to be a robust virial BH mass estimator in  single-epoch scaling relations (\citealp{Greene05,DuP14,Grier17}). In addition, previous studies revealed that the spectral region around H$\beta$ is much less affected by BEL blueshift, intrinsic reddening,  and absorption than \mgii\ or \civ\ (e.g., \citealp{Plotkin15,Coatman19,Yi20}).  Recently, \citet{Yi20}  found observational evidence for BAL winds influencing the \civ\ and \mgii\ BELs more dramatically than the H$\beta$ BEL from a small LoBAL sample at $3\lesssim z\lesssim5$. In agreement with this effect, we find that J0827 shows a striking difference between the \mgii- and H$\beta$-based systemic redshifts ($z=2.04$ vs. $z=2.07$), confirming our initial concern about using the observed \mgii\ BEL as a BH-mass tracer for this quasar (see \citealt{Yi19b}). Therefore, H$\beta$ is the best available tracer of BH mass for J0827.  A general equation of the  single-epoch scaling relation can be expressed as  
\begin{equation}
 {\rm log}  \left[ \frac{ M_{\rm BH}} {M_\odot} \right]  = a +  b\times {\rm log} \left[\frac{\lambda L_{\lambda}} {10^{44}   \rm{erg~s^{-1}}} \right] + 2{\rm log} \left[\frac{\rm FWHM} { \rm{km~s^{-1}}} \right] 
\label{eq1}
\end{equation}

There are several parameter configuration sets ($a,b$) adopted in the literature to estimate the BH mass, depending on the specific calibration. To be consistent with previous WLQ studies (e.g., \citealp{Plotkin15,Luo15,NiQ18}), we use the same prescription ($a=0.7,b=0.5$ for H$\beta$) as  \cite{Plotkin15} to estimate the BH mass for this quasar.   We first perform a local pseudo-continuum fit with a power-law function and an iron template from \citet{Boroson92b}. We then subtract the fitted pseudo continuum from the raw spectrum to obtain the  emission lines of interest. For the emission-line fits, we adopt a similar prescription from the literature (\citealt{Greene05}), in which the velocity separation and dispersion across the narrow emission lines are tied to each other, and the amplitude ratios are fixed to 2.96 for \nii~$\lambda$6583/$\lambda$6548 and \oiii~$\lambda$5007/$\lambda$4959. We do not consider modeling the \sulii~$\lambda$6716/$\lambda$6731 emission due to the apparent lack of such features imprinted on the near-IR spectrum. The broad line region (BLR) component for H$\beta$ or H$\alpha$ is modeled with up to two Gaussians; however, unlike the modeling for narrow emission lines, we do not tie kinematics when modeling the broad H$\beta$ and H$\alpha$ emission.

The simultaneous spectral fit across these emission lines is shown in Fig.~\ref{J0827HaHb_fitting}. The FWHM of the BLR component of H$\beta$ is measured to be 3960$\pm$1090~km~s$^{-1}$ (the uncertainties are measured via a Monte Carlo approach through randomizations of spectral errors in the line-fitting window). The measurements may have larger uncertainties due to telluric absorption on the blue wing, but the total measurement error is still less than the systematic uncertainty of the scaling relation (typically 0.3 dex for H$\beta$) even  considering this uncertainty. Since the above scaling relation has not been calibrated against reverberation mapping samples of BAL quasars, we adopt the systematic uncertainty for conservative purposes. The monochromatic luminosity ($\lambda L_{\lambda}$) at 5100~\AA\ after correcting for Galactic extinction (\citealt{Schlafly11}) is measured to be $\approx 5.89\times 10^{45}$ erg~s$^{-1}$ for J0827.  Since the scaled normal quasar SED is in good agreement with the observed SED at $0.4<\lambda_{\rm rest}<0.8~\mu$m, we choose the unusual reddening curve in \citet{JiangP13} with $E(\lambda-1~\mu {\rm m})\approx0.02$ mag at $\lambda =0.51~\mu$m to correct internal extinction, 
which yields the 5100~\AA\ monochromatic luminosity of $\approx 6.0\times 10^{45}$  erg~s$^{-1}$  for J0827. Note that there is a degeneracy between the reddening and intrinsic SED shape. Combining all these pieces, the BH mass is then estimated to be $M_{\rm BH}=(6.1^{+3.8}_{-2.9} )\times10^8M_\odot$ according to Eq.~\ref{eq1}. This result is also consistent with that derived by H$\alpha$ ($\sim$4.5$\times10^8M_\odot$) using the prescription from \citet{Greene05}. Consequently, the H$\beta$-based Eddington ratio is $\lambda_{\rm Edd}=0.71^{+0.43}_{-0.33}$ with a bolometric correction factor of 9.26 (see \citealt{Yi20} and references therein), suggestive of being in a rapid accretion phase. The  narrow-core component of \oiii\ is weak, perhaps due to shielding associated with dusty outflows (see Section~\ref{discussion_wlq}) 

\subsection{New results from optical spectra}

The optical spectra of J0827 at MJD$\le$ 58,486 were used in our previous studies for this quasar (\citealp{Yi19b,Yi21}). In this work, we have obtained two additional optical spectra from HET/LRS-2 and present all the optical spectra in Fig.~\ref{J0827all_spec_fit_gmos4} for an overview. Following \citet{Yi19b}, all the spectra are normalized using the power-law model to fit the local spectral regions that are visually identified to be free of emission and absorption features. 
For better visual clarity, we smooth all the optical spectra with a 15-pixel ($\sim$900 km~s$^{-1}$) Savitzky-Golay filter window. This smoothing appears optimal for visual inspection.

The Balnicity Index (BI; \citealt{Weymann91}) is widely adopted to search for and characterize BALs. The BI,  whose definition is similar to the rest-frame equivalent width (REW), can be expressed by the following integral:
\begin{equation}
{\rm BI} =  \int _a^b \left[1- \frac{ f(v)} {0.9}\right]  C {\rm d}v 
\label{eq2}
\end{equation}
where $f(v)$is the normalized flux; $C$ is set to zero and only becomes one when the term in the square bracket is continuously positive across a velocity interval of the specific choice, with a flux density less than 90\% of the fitted continuum. $a$ and $b$ are the velocity integration limits, with the canonical values ($a=-3000$ and $b=-25000$ \kms) set to avoid contamination by other ions. However, these limits should be modified in cases where the scientific goals include the investigations of intrinsic narrow absorption lines or extremely high-velocity outflows (e.g., \citealp{Gibson09,Paola20}).  We adopt a velocity interval of 1000 \kms\ to allow the measurements for relatively narrow but intrinsic absorption features. Based on visual inspection of the nine epoch spectra of J0827, $a$ and $b$ are set to be  --40,000 and --10,000 \kms\ for this quasar, within which we identified two  distinct BALs whose flux-weighted centroid velocities are measured at $v_{\rm LOS}=0.098\pm0.003 c$ and $0.075\pm0.002c$ (the uncertainties are obtained from the minimum and maximum centroid velocities over the nine epochs; see the dashed and dotted lines in Fig.~\ref{J0827ew_mjd_seq}), respectively.

\subsubsection{ BAL transformation }

\begin{figure}[h]
\center{} 
 \includegraphics[height=8.0cm,width=8.6cm,  angle=0]{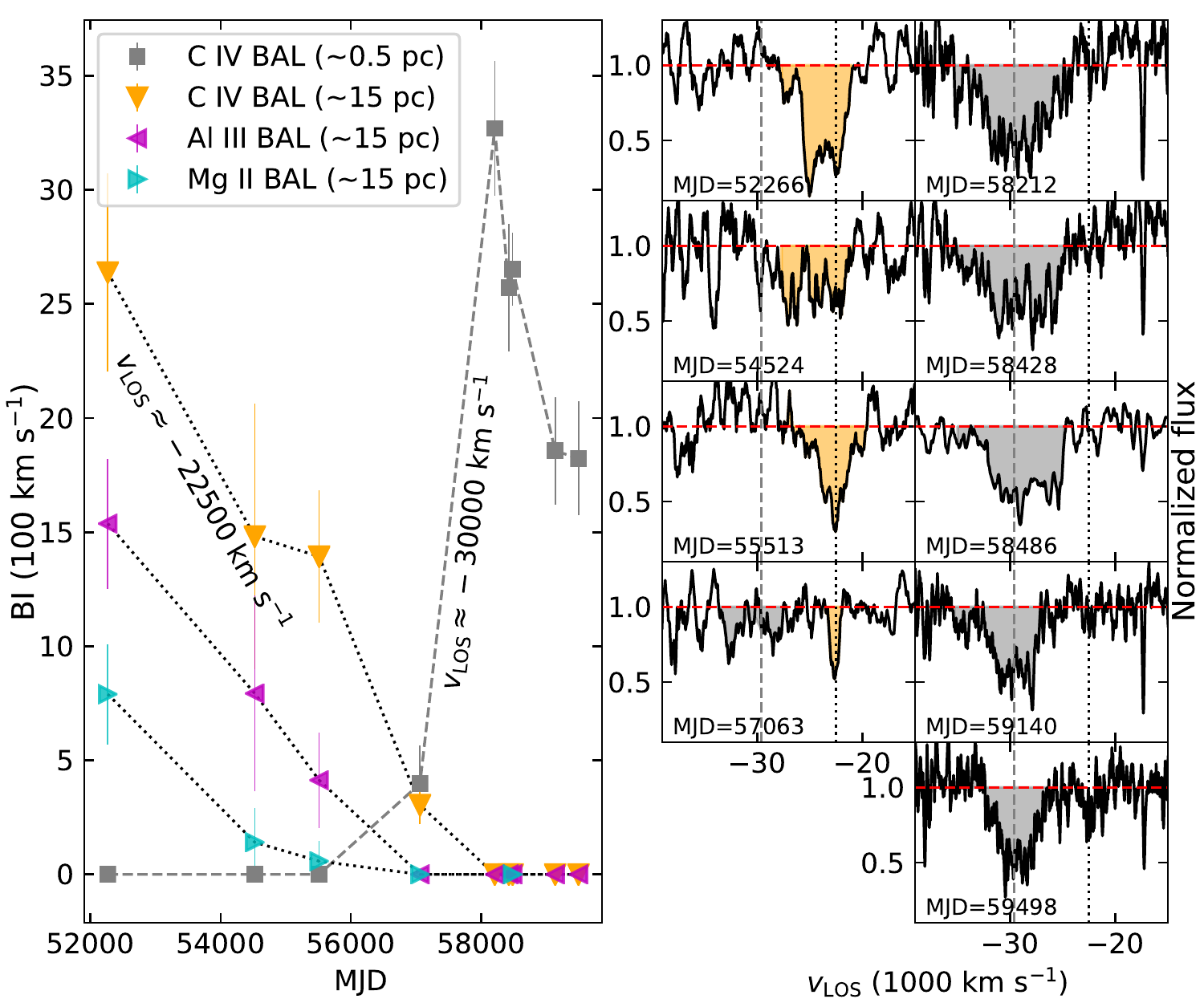} 
      \caption{ Left panel: Time variability of different-velocity, different-ion BALs, in which the \civ\ (gray squares and orange triangles), \aliii\ (magenta triangles), and \mgii\ (cyan triangles) BALs are indicated. Right panel: The corresponding BAL-profile variability over the sampling epochs.  \mgii\ and \aliii\ BAL profiles are not displayed here to avoid overcrowding (see the  color shadings in Fig.~\ref{J0827all_spec_fit_gmos4}). The filled orange/gray colors  depict the BAL disappearance/emergence processes  in time series (left) and velocity space (right). The LoBAL and HiBAL absorbers are located at $\sim$15 and $\sim$0.5 pc from its SMBH (see \citealt{Yi19b}), with the  characteristic velocities marked by the gray dotted and dashed lines. This quasar can serve as an excellent example of the LoBAL$\rightarrow$HiBAL transformation.  }
      \label{J0827ew_mjd_seq}
\end{figure}

 The definition of BI mentioned above is easy to understand,  but one cannot use the BI measurements to determine whether the BALs over different epochs are physically connected. Although the analysis of BAL-complex troughs based on multi-epoch spectra can alleviate this issue to some extent (e.g., \citealp{Filizak13,Yi19a}), ambiguities  cannot be eliminated without sufficient sampling epochs to trace the detailed BAL-profile variability over at least a few rest-frame years.  This is particularly true in the case of J0827, where the BALs varied dramatically from epoch to epoch (see  Fig.~\ref{J0827ew_mjd_seq}); in specific, there are three  BAL-like troughs meeting  BI~$>$~0 at MJD = 57,063, making the assessment of whether they arise from the same region difficult. 
Fortunately, the well-separated sampling epochs of J0827 over two decades, along with the simultaneous BAL disappearance and emergence, provide solid evidence that the gray and orange shaded BALs  are physically distinctive from each other, further in support of transverse motion dominating the BAL disappearance/emergence. Therefore, we  use the two BAL distances ($\sim0.5$ and $\sim15$ pc from the SMBH; see  \citealt{Yi19b})  throughout this work.

As reported in \citet{Yi19b}, one conspicuous feature  is that the low-velocity \civ, \aliii, and \mgii\ BALs at the same LOS velocity slowly weakened until they completely disappeared after MJD = 57,063, then a stronger \civ\ BAL (gray shading) emerges abruptly ($<1$ rest-frame yr) at a higher velocity  (see  Fig.~\ref{J0827ew_mjd_seq}). The simultaneous BAL disappearance/emergence along with such a dramatic velocity shift over less than one rest-frame year robustly rule out BAL acceleration across the entire trough, although acceleration may occur in some BAL subunits, a scenario that cannot be assessed using the data. Our recent two HET/LRS-2 observations confirm the persistence of the newly emerged \civ\ BAL at $v_{\rm LOS}\approx 0.1c$. In addition, there is a large decrease in BAL-trough width from MJD = 58,486 to 59,140, such that the red wing (low-velocity) portion of this BAL trough disappeared (see Fig.~\ref{J0827ew_mjd_seq}); then, the BAL strength appears to level off after MJD = 59,140. The non-zero flux, flat-bottom portion of BALs is indicative of patchy/clumpy outflows partially covering its background light source (e.g., \citealp{Veilleux16,Hamann19}).

A careful examination of Fig.~\ref{J0827all_spec_fit_gmos4} reveals another two notable variability behaviors. First,  the last three-epoch spectra clearly show that the LoBAL species (\aliii\ and \mgii) disappeared and never returned  after MJD = 57,063, confirming the LoBAL$\rightarrow$HiBAL transformation. The lack of \aliii\ BALs seen from the HET/LRS-2 spectra is a strong indicator for the absence of \mgii\ BALs, as both BAL species vary nearly hand-in-hand based on investigations of the LoBAL sample (see Fig.~18 in \citealt{Yi19a}; also see Fig.~\ref{J0827ew_mjd_seq}). Second, the continuum shape becomes progressively bluer in later epochs, suggesting a rapid decrease of dust  along our LOS during the LoBAL$\rightarrow$HiBAL transformation.

We conclude that this quasar provides an excellent example of BAL transformations fitting the evolutionary path from red to blue quasars. Such a transformation may occur many times over the entire quasar lifetime. Longer timescale monitoring is required to confirm whether it is an episodic or a permanent transformation.

\subsubsection{Variability in the  BELs }\label{mgii_bel_var}
\begin{figure}[h]
\center{}
 \includegraphics[height=6cm,width=8.6cm,  angle=0]{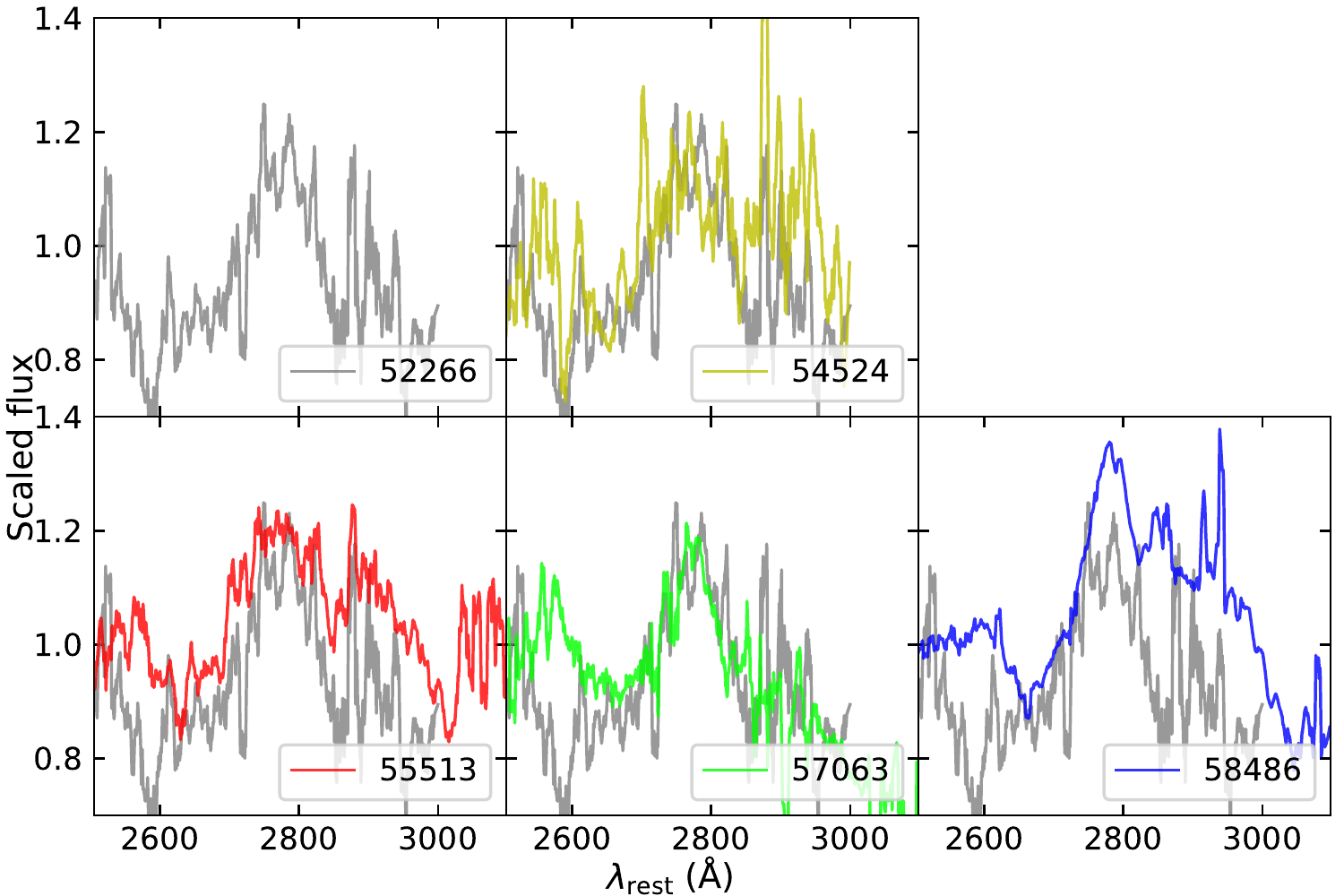} 
  \includegraphics[height=2cm,width=8.6cm,  angle=0]{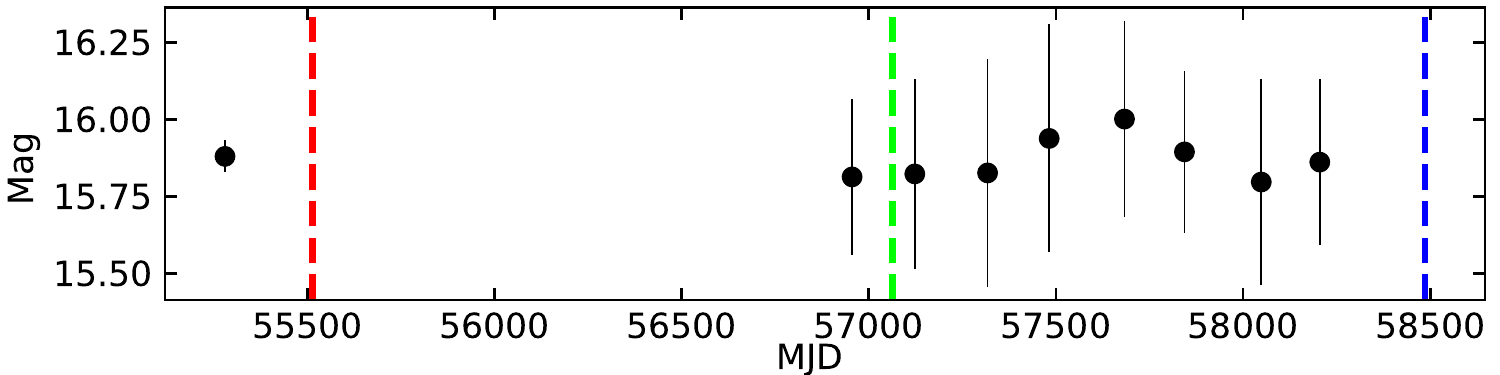} 
      \caption{  Top: The \mgii+\feii\ complex emission over the five different epochs, in which each spectrum is  scaled to an average flux of the relatively line-free region at 2515$<\lambda_{\rm rest}<$2540~\AA, and the first one is used as a benchmark to guide the eye for variability. The complex emission  strengthened substantially from MJD = 57063 to 58486. Bottom: The light curve of W1  shows no significant variability;  dashed lines correspond to the above three spectroscopic epochs. The complex emission strengthened by a factor of $\sim$2 in REW from MJD=52266 to 58486, supportive of an ongoing removal of the dust cocoon.  }
      \label{J0827MgII_profile_var}
\end{figure}
In this subsection, we explore  variability of the  BELs, whose behavior may provide useful diagnostics in assessing subsequent BAL variability after the LoBAL$\rightarrow$HiBAL transformation  and shed light on the nature of WLQs (e.g., \citealp{Diamond09,Plotkin15}).

According to the large difference in systemic redshift  measured by \mgii\ and H$\beta$ (see Section~\ref{systemic_redshift}), the \mgii\ BEL appears to be blueshifted by as much as 2930~\kms\ relative to H$\beta$. However, the previous \mgii-based systemic redshift was determined by the relatively low-S/N SDSS spectra. With the aid of a  higher-quality spectrum from Gemini/GMOS, we are now able to more precisely locate the peak of the \mgii\ BEL at $\lambda_{\rm rest}=2780\pm5$\AA, which corresponds to 2140$\pm$530 \kms\ in blueshift relative to H$\beta$, one of the largest \mgii\ blueshifts known to date (e.g., \citealp{Plotkin15, Luo15,Rafiee16,Venemans16,Vito21}). This value is lower than that derived from the SDSS spectra, partly due to the strengthening in the  red wing of the \mgii+\feii\ complex emission (see Fig.~\ref{J0827MgII_profile_var}). In addition, the \civ\ BEL remains barely visible, consistent with being classified as a bona fide WLQ. The relatively low S/N spectra and the small gap at $1500<\lambda_{\rm rest}<1520$ \AA\ from the HET/LRS-2 spectroscopy make our variability analysis of the \civ\ BEL in J0827 somewhat uncertain (see Fig.~\ref{J0827all_spec_fit_gmos4}). Finally, the Ly$\alpha$+\nv\ emission appears to become  stronger in later epochs, despite the inability to quantify the variability due to its low S/N on the CCD blue edge. Nevertheless, variability of BEL  highlights the importance of a joint analysis of near-IR and  multi-epoch optical spectroscopy for J0827.

Another important issue is the underlying link between the BAL and BEL variability. 
As demonstrated in Fig.~\ref{J0827ew_mjd_seq}, the low-velocity BALs gradually disappeared from MJD = 52,266 to 58,212 and the high-velocity \civ\ BAL emerged rapidly after MJD = 57,063. In contrast, the \mgii\ BEL and the  light curve in the W1 band remains generally unchanged from MJD = 57,063 to 58,212, suggesting  no significant changes in intrinsic quasar luminosity during this time. Due to the lack of W1 data at MJD $>$ 58,212, we cannot  rule out that changes in the incident ionizing flux, a scenario that is hinted by a decrease of the \civ-BAL BI from MJD = 58,486 to 59,140,  plays a (secondary) role in the strengthening of the complex emission from MJD = 57,063 to 58,486, despite transverse motion dominating the BAL transformation. The true situation, however, could be much more complicated given the following reasons: (1) the observed LoBAL disappearance along with a large decrease in continuum reddening strongly supports an ongoing removal of its dust cocoon; (2) BELs and BALs signal the bulk and single-LOS effects, respectively, therefore changes in BAL may not necessarily accompany changes in BEL, particularly when considering patchy/dusty LoBAL winds; (3) the composition and distribution of dust in J0827 remain largely unknown.

\section{ Discussion} \label{discussion_sec}

We now discuss the implications from  joint analyses of the multi-epoch optical spectra sampled over two decades and near-IR spectroscopy, as well as mid-IR photometry for J0827.

\subsection{Implications from the BAL winds} \label{BAL_wind_implication}
Since BAL winds are believed to be launched from a rotating accretion disk  and then accelerated by radiation pressure (e.g. \citealp{Proga00,Grier16}), it is reasonable to assume that the observed LoBAL disappearance is caused by transverse motion with a velocity comparable to the Keplerian velocity. As reported in  \citet{Yi19b}, the LoBAL/HiBAL distances in J0827 were estimated to be $R\sim$15/$\sim$0.5 pc from its SMBH, with roughly an order-of-magnitude uncertainty. The LoBAL distance is consistent with interferometric observations of local AGNs (e.g., \citealp{Honig13}) and the finding from \citet{Ricci17}.  Therefore, the kinetic power ($\dot{E_k}=2\pi R\Omega \mu m_p N_{\rm H}v^3$) for the LoBAL wind is estimated to be $3.45\times10^{46}$ erg~s$^{-1}$ ($\sim$43\% the Eddington luminosity), when adopting a hydrogen column density $N_{\rm H}=10^{22}$~cm$^{-2}$ and a global covering factor $\Omega=0.5$ (both are the lower limits for LoBAL quasars; e.g., \citealp{Boroson92,Gallagher06,Urrutia09,Gibson09,Hamann19}) and a characteristic velocity of 2.25$\times 10^9$ cm~s$^{-1}$ (see Fig.~\ref{J0827ew_mjd_seq}). Such a powerful wind, in principle, is capable of expelling  circumnuclear material  to galactic/circumgalactic scales. Even if the BAL-wind kinetic power  is an order of magnitude lower ($\sim$5\% the Eddington luminosity), its host galaxy could still be  substantially affected by the BAL wind, if provided with a coupling efficiency of $\gtrsim$10\% between the BAL wind and interstellar medium (ISM; see \citealt{Hopkins10}).

Besides the powerful LoBAL wind at $R\sim$15 pc from the SMBH,  there is a HiBAL wind arising at a region whose size ($R\sim$0.5 pc) is comparable to that of the BLR (e.g., \citealp{Grier17} and references therein), which may ultimately clear out the remnant circumnuclear  dust/gas  and speed  the transition of a red quasar  into a blue quasar. This scenario, of course, cannot be directly tested by the optical spectra of J0827 spanning only two decades, but we do see some evidence from its parent sample, which can be summarized as follows: (1) \citet{Yi19a} discovered that weakening BALs outnumber strengthening BALs  on longer sampling timescales in the LoBAL sample; (2) \citet{Yi21} further revealed that the remarkable time-dependent variability in LoBALs described above occurs in the regime where quasars become bluer in later epochs (see Fig.~4 in that work); and (3) all six LoBAL quasars undergoing LoBAL$\rightarrow$HiBAL/non-BAL transformations in \citet{Yi21} became bluer in later epochs. Together with the broad range of striking properties observed in J0827, we suggest that LoBAL quasars such as J0827 could be caught in the act of casting off their dust cocoons, given that LoBALs have high column densities and are almost certainly associated with dust, and that the inclination is likely intermediate or possibly closer to its polar axis due to the type-1 quasar nature.

There are interesting implications of the two different-scale BAL winds located at $R\sim0.5$ and $R\sim15$ pc from the quasar center.  Recent observational studies based on large samples suggest that the majority of BAL winds are distributed in a wide distance range, i.e., at \break $R\sim$10--1000 pc from their SMBHs (e.g., \citealp{Arav18, He19}). Such a wide distribution of BAL-wind scales can be better understood in the evolutionary path from red to blue quasars, given that J0827 is a LoBAL quasar surrounded by a heavy dust cocoon while these samples mainly consist of HiBAL quasars, whose dust cocoons have been mostly dispersed by previous winds. This path is supported by (1) the overall LoBAL$\rightarrow$HiBAL transformation sequence along with a decrease in reddening observed from the LoBAL sample (\citealp{Yi21}); and (2) the larger-distance BAL wind having a lower LOS velocity observed  in J0827 (see Fig.~\ref{J0827ew_mjd_seq}), although we did not detect evident acceleration  signatures in this quasar. However, non-detection of BAL-acceleration signatures over the nine epochs could be due to the saturation effect indicated by non-zero flux, flat-bottom portions of the BALs, the blending of numerous subunits across a BAL trough, the short time span (seven rest-frame yr) compared to a typical BAL lifetime of at least hundreds of years (e.g., \citealp{Filizak13,Yi21}), and/or the insufficient spectral quality.

Indeed, we find evidence of BAL acceleration in a few brighter  quasars accompanying a conspicuous decrease or even complete disappearance of LoBAL species (Yi et al., in prep). Observations of these quasars may offer valuable  clues for the open, extremely challenging question---how quasar winds travel from nuclear  to galactic scales? Given the near Eddington accretion and dusty/patchy LoBAL wind observed in J0827, radiation pressure on dust likely plays an important role in the feedback of momentum and energy from nuclear to galactic scales for this process, in agreement with the predications from some theoretical works (e.g., \citealp{Fabian08,Costa14, Giustini19}) and observational studies (e.g., \citealp{Ricci17,Coatman19,Leftley19,Yi20}). 
We also aim to explore the BAL-ISM coupling process from an excellent later-phase analog of J0827, another brighter/bluer WLQ at $z\sim2$ showing BAL deceleration, weak radio emission at 1.4 GHz,  excess emission at $\lambda_{\rm rest}=7.5~\mu$m, and multip-scale/multi-phase outflows, in a future study. Such investigations may help to reveal the origin of quasar reddening and shed light on the lack of significant differences in emission-line outflows between red and blue quasars (see below).

\subsection{Implications from the SED } \label{discussion_sed}

  The most dramatic feature  seen from Fig.~\ref{J0827sed_nonBAL2021Nov} is that the observed SED of J0827 is in good agreement with the scaled normal quasar SED only at $\lambda_{\rm rest}>0.4~\mu$m but deviates increasingly at $\lambda_{\rm rest}<0.3~\mu$m. Although the SMC extinction curve has been  widely adopted in the literature for studies of quasar dust properties, it cannot explain such a behavior in J0827, unless with a break at $\lambda_{\rm rest}\sim0.3~\mu$m. In contrast, the anomalous extinction curve from \citet{JiangP13} can satisfactorily reproduce this behavior without the need to invoke two  extinction curves at $\lambda_{\rm rest}\lesssim0.3~\mu$m and $\lambda_{\rm rest}\gtrsim0.3~\mu$m. From a physical perspective,  i.e., quasar activities can cause the destruction of large dust grains or in situ formation of dust grains may occur in quasar winds, the extinction curve from \citet{JiangP13} is  an optimal choice to fit the entire SED for J0827, provided that its intrinsic SED is closely similar to the  normal quasar SED. 

A number of observational studies have reported significant correlations between \civ\ BEL/BAL properties (i.e. blueshift and EW) and  dust  properties traced by rest-frame 2--6~$\mu$m emission and UV/NIR spectral slopes  (e.g., \citealp{Zhang14, Baron19,Rankine20, Yi20, Temple21}), suggestive of an underlying link between quasar wind and dust across 1--1000~pc scales. However, these studies suffered from  large uncertainties in distance for both wind and dust, making robust assessments difficult. 
For example, \citet{Yi20} found significant differences in the  narrow line region (NLR) and BLR kinematics between the BAL and non-BAL samples matched in redshift and luminosity, such that BAL quasars tend to have larger  \civ\ and \oiii\ blueshifts. 
Similarly,  \citet{Rivera21} discovered a higher incidence of outflow signatures traced by large \civ-BEL blueshifts and  broad \oiii\ wings  in red quasars compared to their control sample, implicating pc- and kpc-scale dusty winds as a potential key ingredient for red quasars. 
In contrast, some  studies found no significant differences in outflow properties between red and blue quasars matched in luminosity (e.g., \citealp{Temple19,Villar-Martin20,Fawcett22}). The apparent discrepancy could be due primarily to the difference in sample size  and/or  the specific fitting methods, such as the difficulties in decomposing robust \oiii\ kinematics from type-1 quasar spectra (e.g., \citealp{Coatman19,Villar-Martin20}). Therefore, unambiguous quasar winds like BALs, particularly those with a good constraint on the wind distance, provide   valuable diagnostics to study the underlying link between wind and dust.

Fortunately, the remarkable BAL transitions of J0827 allow us to constrain its LoBAL wind distance at $R\sim$15~pc from the quasar center (see \citealp{Yi19b}). In combination with the estimated kinetic power ($\sim$43\% the Eddington luminosity) of the LoBAL wind, the strengthening in \mgii\ emission suggestive of dust removing, and a large decrease of reddening after the LoBAL$\rightarrow$HiBAL transformation, it is likely that this quasar is caught in the act of casting off its dust cocoon located at $R\sim$15 pc, with an  outermost shell perhaps extending to $R\sim$1000~pc.   We conclude that J0827 offers a clear example for dusty winds as the origin of quasar reddening, which, in turn, supports the excess emission at $\lambda_{\rm rest}=7.0$ $\mu$m (e.g., \citealp{Baron19,Rivera21}).

\subsection{ A potential link between WLQs and BAL quasars }  \label{discussion_wlq}

The majority of WLQs known to date have been identified from single-epoch spectroscopy; in addition, the current WLQ samples targeted at X-ray wavelengths exclude quasars with apparent absorption features (e.g., \citealp{WuJ12,Luo15,NiQ18}), making the link to outflows unclear. While \citet{Plotkin15} proposed a ``wind-dominated'' scenario for typical non-BAL WLQs on the basis of their BEL properties, there is some debate if winds are traced by emission lines (e.g., \citealp{Gaskell82,Richards11,Zuo20,Villar-Martin20}). Recently, \citet{NiQ18} invoked a thick disk+outflow model for WLQs, but they implicitly considered both components as the shield material and did not consider the impact of disk winds to the circumnuclear or  larger-scale environment. 
Therefore, the connection between WLQs and outflow properties remains largely unexplored in these studies. In this context, searching for unambiguous quasar winds like BALs among WLQs is particularly valuable.

\citet{Rogerson16} reported a dramatic non-BAL$\rightarrow$HiBAL transformation in a quasar J0230, which was considered a WLQ due to the persistence of extremely weak \civ\ emission over one decade. Similarly,  from our ongoing spectroscopic  campaign on quasars showing BAL transitions, we discovered a more surprising phenomenon: a WLQ is undergoing multiple HiBAL$\rightleftharpoons$non-BAL transformations over two decades (Yi et al., in prep). In addition,  \citet{JiangP13} demonstrated that the \civ\ BEL of their LoBAL quasar remains weak after the correction for internal reddening (see Fig.~5 in their work).  Likewise, the \civ\ BEL in J0827 is much weaker than that from the quasar composite reddened by the same reddening curve from \citet{JiangP13} to match the continuum level of J0827 (see Fig.~\ref{J0827all_spec_fit_gmos4}). Moreover, J0827 is undergoing a LoBAL$\rightarrow$HiBAL transformation and its \civ\ BEL appears barely visible over all the spectroscopic epochs; in particular, the optical spectrum at MJD = 57,063 did not show conventional BAL features and remained barely visible in the \civ\ BEL, making it a bona fide WLQ. 
The nearest quasar, Mrk 231 (\citealt{Veilleux16}), resembles J0827 in many aspects, such as weak high-ionization BELs compared to normal H$\alpha$ emission, LoBAL winds at $R\sim$2--20 pc, large BEL blueshift, weak or absent \oiii\ emission, near Eddington-ratio accretion, a sudden spectral turning point at $\lambda_{\rm rest}\sim0.3~\mu$m, a  peak of the SED at $\lambda_{\rm rest}\sim 7~\mu$m, and a time-variability pattern characterized by a rapid rise and a slow decline in BAL strength (\citealp{Lipari05}). Intriguingly, \citet{Leighly14} reported tentative evidence for BAL deceleration in Mrk 231, a telltale sign for the BAL-ISM coupling in action. 
These results indicate WLQs having BAL and dust features are an important population to bridge the gap between red and blue WLQs, presumably because BAL winds play a role in shedding  dust cocoons (see Section~\ref{BAL_wind_implication}). In agreement with \citet{Yi21}, the above five WLQs, among which three are LoBAL quasars that are much redder than the other two HiBAL quasars, together fit the predominant LoBAL$\rightarrow$HiBAL/non-BAL transformation sequence along with a decease in the amount of dust, further supporting the evolutionary path of red quasars transitioning into blue quasars. Weak, narrow absorption lines (NALs) may be more prevalent in WLQs than previously thought (Paul et al., submitted), but we do not discuss this issue due to the difficulty of identifying quasar outflows traced by NALs.

Almost all  WLQ studies have not explored the time variability of BELs. Through visual inspection of objects with multi-epoch SDSS spectra in two previous WLQ studies, we found that the ``pristine'' WLQs from \citet{Luo15} vary more frequently and dramatically than the ``bridge'' population of WLQs from \citet{NiQ18} with respect to emission, absorption, and/or continuum properties, i.e., 4 out of 17 quasars with multi-epoch spectra from \citet{Luo15} show appreciable variability of the \civ\ BEL in REW as opposed to non such cases in the bridge quasars from \citet{NiQ18}. In contrast, the multi-epoch optical spectroscopy of J0827 reveals that both the \mgii\ and \civ\ BELs strengthened along with a decrease of the \mgii\ blueshift in later epochs.  Thus, our initial investigation of variability in BEL between WLQs and bridge quasars suggests a one-way direction, such that bona fide WLQs transition into the bridge population (similar to the LoBAL$\rightarrow$HiBAL/non-BAL sequence mentioned above), hinting that WLQs may be at an earlier phase than the bridge quasars. However, we caution about a potential (though unlikely) bias due to the lack of a thorough examination for normal quasars transitioning into WLQs. Such a transition, to our knowledge, has not been reported but might occur, i.e., \citet{Ross20} found 3 out of 64774 quasars exhibiting  dramatic changes in the \civ\ BEL, although none of them has \civ\ REWs less than 15 \AA\ at any sampling epochs. 
 Nevertheless, the persistence of large \mgii\ blueshift (see Section~\ref{mgii_bel_var}) and weak \civ\ emission (see Fig.~\ref{J0827all_spec_fit_gmos4}) in J0827 suggests that this quasar may still have to spend a considerable time (perhaps a few million yr) before evolving into a normal, blue quasar.

Our spectral fitting for J0827 reveals a rapid accretion state ($\lambda_{\rm Edd}\sim0.71$),  supporting the argument that WLQs may be in an exceptionally high accreting phase, i.e., near- or super-Eddington accretors (e.g., \citealp{Luo15,Veilleux16,NiQ18}). This interpretation is supported by the potential  excess  emission at $\lambda_{\rm rest} \approx 7\mu$m as justified in Section~\ref{discussion_sed}, which may lead to an underestimation of the bolometric luminosity, although the excess could be contributed predominantly by dusty outflows (e.g., \citealp{Honig13,Zakamska14,Baron19}). Combined with all the H$\beta$-based Eddington ratios larger than 0.3 found in a non-BAL WLQ sample from \citet{Plotkin15}, makes it tempting to conclude that WLQs have near or super-Eddington ratios with accompanying thick disks (e.g., \citealp{WangJ14,Luo15,Veilleux16,NiQ18,Giustini19}). Likewise, high-$z$ BAL quasars tend to possess near or super-Eddington ratios estimated by H$\beta$ and \mgii\ (e.g., \citealp{WangF18,Yi20}). Therefore, high Eddington ratio ($\gtrsim$0.3) may be a fundamental quantity in connecting both the WLQ and BAL populations.  We briefly outline this argument below.

Previous studies of WLQs, mainly from an X-ray perspective, suggest a ``shielding'' scenario (e.g., \citealp{WuJ12,Luo15,NiQ18}). Similarly, some  BAL models also require a ``shielding'' mechanism by which high-energy photons are blocked from reaching the UV absorbers (e.g., \citealp{Proga00}). From an observational view, both WLQs and BAL quasars with weak BELs favor a soft ionizing SED (e.g., \citealp{Plotkin15,Hamann19,Rankine20}; Paul et al., submitted), consistent with the prediction from shielding.  Such shielding would naturally produce high-velocity outflows, which is supported by the prevalence of large \civ\ BEL blueshifts seen in WLQs and subrelativistic winds observed in BAL quasars (e.g., \citealp{Luo15,Yi19a,Rankine20}). Additionally, weak narrow-core \oiii\ emission is often seen in LoBAL quasars and WLQs (e.g., \citealp{Boroson92b,WuJ11,Plotkin15,Yi20}), in agreement with shielding. Indeed, shielding lends  support to the argument that BEL blueshift, BAL, and perhaps broad \oiii\ emission could be different manifestations of the same quasar-driven outflow system (e.g., \citealp{Rankine20,Xu20,Yi20,LiuB21}), presumably because shielding plays a key role in producing powerful disk winds that can travel out to large scales. In contrast, the LoBAL wind and weak BELs/narrow-core \oiii\ emission are indicative of shielding. Therefore, in the context of shielding resulting from high Eddington accretion, the WLQ, BAL, and weak narrow-core \oiii\ emission may be unified  for J0827. However,  realistic shielding could be much more complicated than the discussion here (e.g., \citealp{Nenkova08,Ricci17,NiQ20}), as the nature of the shielding material remains unknown.

\section{Summary and future work}\label{disc_con}

The main results and implications from  joint analyses of the multi-wavelength  data are concluded as follows.

 \begin{enumerate}
\item
The systemic redshift of J0827 is measured to be $z=2.070\pm0.001$ using H$\beta$, significantly larger than the \mgii-based one ($z=2.040\pm0.003$) reported from the literature. Our measurements yield an extreme \mgii\ blueshift of 2140$\pm$530 \kms\ relative to H$\beta$ (see Section~\ref{systemic_redshift}).
\item
Based upon the single-epoch scaling relation, the BH mass and Eddington ratio are estimated to be $M_{\rm BH}=6.1\times10^8~M_\odot$ and $\lambda_{\rm Edd}=0.71$, with a typical uncertainty of 0.3 dex (see Section~\ref{BH_mass}). 
\item
The LoBAL wind is thought to be located at $R\sim$15 pc from its BH. Combined with the type-1 quasar nature, J0827 fits better to a dust cocoon than the conventional torus. Adopting typical values of the global covering factor (0.5) and hydrogen column density ($10^{22}$ cm$^{-2}$) in LoBAL quasars, the wind kinetic power is estimated to be 43\% the Eddington luminosity with an order-of-magnitude uncertainty. J0827 provides an excellent laboratory to test the evolutionary path from red to blue quasars, given the (episodic) LoBAL$\rightarrow$HiBAL transformation along with a rapid decrease in reddening characteristic of the blowout phase (see Section~\ref{BAL_wind_implication}). 
\item
Our analyses support the presence of excess emission at rest-frame $7~\mu$m for J0827 compared to normal quasars, on the basis of dusty outflows traced by the powerful LoBAL wind (see Section~\ref{discussion_sed}). 
\item
Based upon visual inspection of variability among the five WLQs with BALs, typical WLQs without BALs, and bridge quasars,  WLQs tend to vary more frequently and dramatically than the bridge quasars in emission, absorption, and continuum (see Section~\ref{discussion_wlq}). 
\item
Our investigations of J0827 complement the ``wind-dominated'' scenario invoked for WLQs and suggest that both WLQs and BAL (particularly LoBAL) quasars favor soft ionizing SEDs and high-Eddington ratios, and that the WLQ, BAL, and weak narrow-core \oiii\ emission  phenomena may be unified in the broad context of shielding for this quasar (see Section~\ref{discussion_wlq}).

\end{enumerate}

It is well established that LoBAL quasars are considerably redder than HiBAL/non-BAL quasars, and that BAL fraction is much higher in red quasars than in blue quasars. While leading models of galaxy formation predict an evolutionary path from red to blue quasars,  solid observational evidence remains lacking, perhaps due to the short-lived nature of red quasars. As a case study, the observed LoBAL$\rightarrow$HiBAL transformation in J0827, along with the LoBAL wind at $R\sim$15 pc  and a large decrease of reddening over two decades, provides a unique opportunity to demonstrate that BAL winds play a critical role in transforming red quasars into blue quasars via the rapid removal of their dust cocoons. Therefore,  J0827 can serve as an excellent example of BAL winds destroying dust/gas cocoons, a scenario that has  long been invoked by observational and theoretical studies.  Furthermore, we will test this scenario in detail in combination with other quasars showing BAL transformations and acceleration signatures in the future.

In addition, longer time baseline and wider wavelength coverage observations of J0827 can  trace its time evolution, constrain the star-formation rate, explore the radio emission, study the dust properties, and further assess whether this quasar is at the end of the blowout phase toward unveiling a blue quasar, or just reflects an episodic ignition of quasar activities from a dusty environment.

\acknowledgments
We thank the anonymous referee for constructive feedback that helped improve this manuscript. 
We are grateful to Robin Ciardullo and Michael Eracleous for assistance with the observations by the Hobby-Eberly Telescope. 
W.Yi, WNB and QN acknowledge support from  NSF grant AST-2106990, CXC grant GO0-21080X, and the V.M. Willaman Endowment at Penn State.   
W.Yi also thanks support from the National Science Foundation of China (11703076) and the West Light Foundation of The Chinese Academy of Sciences (Y6XB016001). 
LCH was supported by the National Science Foundation of China (11721303, 11991052) and the National Key R\&D Program of China (2016YFA0400702). B.L. acknowledges financial support from the National Natural Science Foundation of China grant 11991053. W.Yi, X.-B. Wu and J.-M. Bai acknowledge the science research grants from the China Manned Space Project with No. CMS-CSST-2021-A06. J.-M. Bai acknowledges the National Natural Science Foundation of China grant 11991051.

This work uses data obtained from the Gemini Observatory (PI: Yi; program ID:  GN-2018B-FT-214 and GN-2020A-FT-203), which is operated by the Association of Universities for Research in Astronomy, Inc., under a cooperative agreement with the NSF on behalf of the Gemini partnership: the National Science Foundation (United States), the National Research Council (Canada), CON- ICYT (Chile), Ministerio da Ciencia, Tecnologia e Inovaciao (Brazil) and Ministerio de Ciencia, Tecnologia e Innovacion Productiva (Argentina). 
This research also uses data obtained through the Telescope Access Program (TAP), which has been funded by the National Astronomical Observatories of China, the Chinese Academy of Sciences (the Strategic Priority Research Program ''The Emergence of Cosmological Structures'' grant No. XDB09000000), and the Special Fund for Astronomy from the Ministry of Finance. Observations obtained with the Hale Telescope at Palomar Observatory were obtained as part of an agreement between the National Astronomical Observatories, the Chinese Academy of Sciences, and the California Institute of Technology. 
The Hobby-Eberly Telescope (HET) is a joint project of the University of Texas at Austin, the Pennsylvania State University, Ludwig-Maximillians-Universit\"{a}t M\"{u}nchen, and Georg-August-Universit\"{a}t G\"{o}ttingen. The Hobby-Eberly Telescope is named in honour of its principal benefactors, William P. Hobby and Robert E. Eberly. 
The Low-Resolution Spectrograph 2 (LRS2) was developed and funded by the
University of Texas at Austin McDonald Observatory and Department of
Astronomy, and by the Pennsylvania State University. We thank the
Leibniz-Institut f\"ur Astrophysik Potsdam and the Institut f\"ur
Astrophysik G\"ottingen for their contributions to the construction
of the integral field units. 
Funding for SDSS-III has been provided by the Alfred P. Sloan Foundation, the Participating Institutions, the National Science Foundation, and the U.S. Department of Energy Office of Science.

\newpage


\begin{thebibliography}{}
\small


\bibitem[Allen et al.(2011)]{Allen11} Allen, J.~T., Hewett, P.~C., Maddox, N., Richards, G.~T., \& Belokurov, V.\ 2011, \mnras, 410, 860
\bibitem[Arav et al.(2018)]{Arav18}Arav, N., Liu, G., Xu, X., et al. 2018, \apj, 857, 60 
\bibitem[Baron \& Netzer(2019)]{Baron19} Baron, D. \& Netzer, H.\ 2019, \mnras, 482, 3915
\bibitem[Banerji et al.(2015)]{Banerji15} Banerji, M., Alaghband-Zadeh, S., Hewett, P.~C., et al.\ 2015, \mnras, 447, 3368
\bibitem[Boroson \& Meyers(1992)]{Boroson92} Boroson, T.~A. \& Meyers, K.~A.\ 1992, \apj, 397, 442
\bibitem[Boroson \& Green(1992)]{Boroson92b} Boroson, T.~A. \& Green, R.~F.\ 1992, \apjs, 80, 109
\bibitem[Chonis et al.(2014)]{Chonis14}   Chonis, T. S. et al., 2014, SPIE, 9147
\bibitem[Calistro Rivera et al.(2021)]{Rivera21} Calistro Rivera, G., Alexander, D.~M., Rosario, D.~J., et al.\ 2021, \aap, 649, A102
\bibitem[Coatman et al.(2019)]{Coatman19} Coatman, L., Hewett, P.~C., Banerji, M., et al.\ 2019, \mnras, 486, 5335
\bibitem[Costa et al.(2014)]{Costa14} Costa, T., Sijacki, D., \& Haehnelt, M.~G.\ 2014, \mnras, 444, 2355
\bibitem[Davis et al.(2018)]{Davis18} Davis, B.~D., Ciardullo, R., Jacoby, G.~H., et al.\ 2018, \apj, 863, 189
\bibitem[Diamond-Stanic et al.(2009)]{Diamond09} Diamond-Stanic, A.~M., Fan, X., Brandt, W.~N., et al.\ 2009, \apj, 699, 782
\bibitem[De Cicco et al.(2018)]{DeCicco18} De Cicco, D., Brandt, W.~N., Grier, C.~J., et al.\ 2018, \aap, 616, A114
\bibitem[Du et al.(2014)]{DuP14} Du, P., Hu, C., Lu, K.-X., et al.\ 2014, \apj, 782, 45782/1/45
\bibitem[Fabian et al.(2008)]{Fabian08} Fabian, A.~C., Vasudevan, R.~V., \& Gandhi, P.\ 2008, \mnras, 385, L43
\bibitem[Fan et al.(1999)]{Fan99} Fan, X., Strauss, M.~A., Gunn, J.~E., et al.\ 1999, \apjl, 526, L57
\bibitem[Filiz Ak et al.(2013)]{Filizak13} Filiz Ak, N., Brandt, W.~N., Hall, P.~B., et al.\ 2013, \apj, 777, 168 
\bibitem[Fawcett et al.(2022)]{Fawcett22} Fawcett, V.~A., Alexander, D.~M., Rosario, D.~J., et al.\ 2022, arXiv:2201.04139
\bibitem[Fynbo et al.(2013)]{Fynbo13} Fynbo, J.~P.~U., Krogager, J.-K., Venemans, B., et al.\ 2013, \apjs, 204, 6
\bibitem[Gaskell(1982)]{Gaskell82} Gaskell, C.~M.\ 1982, \apj, 263, 79
\bibitem[Gallagher et al.(2006)]{Gallagher06} Gallagher, S.~C., Brandt, W.~N., Chartas, G., et al.\ 2006, \apj, 644, 709 
\bibitem[Gibson et al.(2009)]{Gibson09} Gibson, R.~R., Jiang, L., Brandt, W.~N., et al.\ 2009, \apj, 692, 758 
\bibitem[Giustini \& Proga(2019)]{Giustini19} Giustini, M. \& Proga, D.\ 2019, \aap, 630, A94
\bibitem[Glikman et al.(2006)]{Glikman06} Glikman, E., Helfand, D.~J., \& White, R.~L.\ 2006, \apj, 640, 579
\bibitem[Glikman et al.(2012)]{Glikman12} Glikman, E., Urrutia, T., Lacy, M., et al.\ 2012, \apj, 757, 51
\bibitem[Greene \& Ho(2005)]{Greene05} Greene, J.~E. \& Ho, L.~C.\ 2005, \apj, 630, 122
\bibitem[Grier et al.(2016)]{Grier16} Grier, C.~J., Brandt, W.~N., Hall, P.~B., et al.\ 2016, \apj, 824, 130
\bibitem[Grier et al.(2017)]{Grier17} Grier, C.~J., Trump, J.~R., Shen, Y., et al.\ 2017, \apj, 851, 21
\bibitem[Hamann et al.(2019)]{Hamann19}Hamann, F.; Herbst, H.; Paris, I. \& Capellupo, D. 2019, \mnras, 483, 1808
\bibitem[He et al.(2019)]{He19} He, Z., Wang, T., Liu, G., et al.\ 2019, Nature Astronomy, 3, 265
\bibitem[Hill et al.(2021)]{Hill21} Hill, G.~J., Lee, H., MacQueen, P.~J., et al.\ 2021, arXiv:2110.03843
\bibitem[H{\"o}nig et al.(2013)]{Honig13} H{\"o}nig, S.~F., Kishimoto, M., Tristram, K.~R.~W., et al.\ 2013, \apj, 771, 87
\bibitem[Hopkins \& Elvis(2010)]{Hopkins10} Hopkins, P.~F., \& Elvis, M.\ 2010, \mnras, 401, 7
\bibitem[Indahl et al.(2019)]{Indahl19} Indahl, B.; Zeimann, G.; Hill, G. J. Et al. 2019, \apj, 883, 114  
\bibitem[Jiang et al.(2013)]{JiangP13} Jiang, P., Zhou, H., Ji, T., et al.\ 2013, \aj, 145, 157
\bibitem[Leftley et al.(2019)]{Leftley19} Leftley, J.~H., H{\"o}nig, S.~F., Asmus, D., et al.\ 2019, \apj, 886, 55
\bibitem[Leighly et al.(2014)]{Leighly14} Leighly, K.~M., Terndrup, D.~M., Baron, E., et al.\ 2014, \apj, 788, 123
\bibitem[L{\'\i}pari et al.(2005)]{Lipari05} L{\'\i}pari, S., Terlevich, R., Zheng, W., et al.\ 2005, \mnras, 360, 416
\bibitem[Luo et al.(2015)]{Luo15} Luo, B. et al. 2015, \apj, 805, 122
\bibitem[Liu et al.(2021)]{LiuB21} Liu, B., Zhou, H.-Y., Shu, X.-W., et al.\ 2021, Research in Astronomy and Astrophysics, 21, 065
\bibitem[Nenkova et al.(2008)]{Nenkova08} Nenkova, M., Sirocky, M.~M., Nikutta, R., et al.\ 2008, \apj, 685, 160
\bibitem[Ni et al.(2018)]{NiQ18} Ni, Q., Brandt, W.~N., Luo, B., et al.\ 2018, \mnras, 480, 5184
\bibitem[Ni et al.(2020)]{NiQ20} Ni, Q., Brandt, W.~N., Yi, W., et al.\ 2020, \apjl, 889, L37
\bibitem[Plotkin et al.(2015)]{Plotkin15} Plotkin, R.~M., Shemmer, O., Trakhtenbrot, B., et al.\ 2015, \apj, 805, 123
\bibitem[Proga et al.(2000)]{Proga00} Proga, D., Stone, J.~M., \& Kallman, T.~R.\ 2000, \apj, 543, 686 
\bibitem[Rogerson et al.(2016)]{Rogerson16} Rogerson, J.~A., Hall, P.~B., Rodr{\'\i}guez Hidalgo, P., et al.\ 2016, \mnras, 457, 405
\bibitem[Rogerson et al.(2018)]{Rogerson18} Rogerson, J.~A., Hall, P.~B., Ahmed, N.~S., et al.\ 2018, \apj, 862, 22 
\bibitem[Rodr{\'\i}guez Hidalgo et al.(2020)]{Paola20} Rodr{\'\i}guez Hidalgo, P., Khatri, A.~M., Hall, P.~B., et al.\ 2020, \apj, 896, 151    
\bibitem[Rafiee et al.(2016)]{Rafiee16} Rafiee, A., Pirkola, P., Hall, P.~B., et al.\ 2016, \mnras, 459, 2472  
\bibitem[Ramsey et al.(1998)]{Ramsey98} Ramsey, L.~W., Adams, M.~T., Barnes, T.~G., et al.\ 1998, \procspie, 3352, 34
\bibitem[Rankine et al.(2020)]{Rankine20} Rankine, A.~L., Hewett, P.~C., Banerji, M., et al.\ 2020, \mnras, 492, 4553
\bibitem[Ricci et al.(2017)]{Ricci17} Ricci, C., Trakhtenbrot, B., Koss, M.~J., et al.\ 2017, \nat, 549, 488
\bibitem[Richards et al.(2006)]{Richards06} Richards, G.~T., Lacy, M., Storrie-Lombardi, L.~J., et al.\ 2006, \apjs, 166, 470      
\bibitem[Richards et al.(2011)]{Richards11} Richards, G.~T., Kruczek, N.~E., Gallagher, S.~C., et al.\ 2011, \aj, 141, 167
\bibitem[Richards et al.(2003)]{Richards03} Richards, G.~T., Hall, P.~B., Vanden Berk, D.~E., et al.\ 2003, \aj, 126, 1131
\bibitem[Rankine et al.(2020)]{Rankine20} Rankine, A.~L., Hewett, P.~C., Banerji, M., et al.\ 2020, \mnras, 492, 4553
\bibitem[Ross et al.(2020)]{Ross20} Ross, N.~P., Graham, M.~J., Calderone, G., et al.\ 2020, \mnras, 498, 2339 
\bibitem[Schlafly \& Finkbeiner(2011)]{Schlafly11} Schlafly, E.~F. \& Finkbeiner, D.~P.\ 2011, \apj, 737, 103
\bibitem[Temple et al.(2019)]{Temple19} Temple, M.~J., Banerji, M., Hewett, P.~C., et al.\ 2019, \mnras, 487, 2594 
\bibitem[Temple et al.(2021)]{Temple21} Temple, M.~J., Banerji, M., Hewett, P.~C., et al.\ 2021, \mnras, 501, 3061
\bibitem[Timlin et al.(2020)]{Timlin20} Timlin, J.~D., Brandt, W.~N., Ni, Q., et al.\ 2020, \mnras, 492, 719
\bibitem[Urrutia et al.(2009)]{Urrutia09} Urrutia, T., Becker, R.~H., White, R.~L., et al.\ 2009, \apj, 698, 1095
\bibitem[Voit et al.(1993)]{Voit93} Voit, G.~M., Weymann, R.~J., \& Korista, K.~T.\ 1993, \apj, 413, 95
\bibitem[Veilleux et al.(2016)]{Veilleux16} Veilleux, S., Mel{\'e}ndez, M., Tripp, T.~M., et al.\ 2016, \apj, 825, 42
\bibitem[Venemans et al.(2016)]{Venemans16} Venemans, B.~P., Walter, F., Zschaechner, L., et al.\ 2016, \apj, 816, 37
\bibitem[Vito et al.(2021)]{Vito21} Vito, F., Brandt, W.~N., Ricci, F., et al.\ 2021, \aap, 649, A133
\bibitem[Villar Mart{\'\i}n et al.(2020)]{Villar-Martin20} Villar Mart{\'\i}n, M., Perna, M., Humphrey, A., et al.\ 2020, \aap, 634, A116 
\bibitem[Vanden Berk et al.(2001)]{VandenBerk01}Vanden Berk, D. E. et al. 2001, \aj, 122, 549
\bibitem[Wang et al.(2014)]{WangJ14} Wang, J.-M., Qiu, J., Du, P., et al.\ 2014, \apj, 797, 65
\bibitem[Wang et al.(2015)]{WangT15} Wang, T., Yang, C., Wang, H., et al.\ 2015, \apj, 814, 150
\bibitem[Wang et al.(2018)]{WangF18} Wang, F., Yang, J., Fan, X., et al.\ 2018, \apjl, 869, L9
\bibitem[Wright et al.(2010)] {Wright2010} Wright, E. L., Eisenhardt, P. R. M., Mainzer, A. K., et al. 2010, \aj, 140, 1868
\bibitem[Weymann et al.(1991)]{Weymann91} Weymann, R.~J., Morris, S.~L., Foltz, C.~B., et al.\ 1991, \apj, 373, 23
\bibitem[Wilson et al.(2004)]{Wilson04} Wilson, J.~C., Henderson, C.~P., Herter, T.~L., et al.\ 2004, \procspie, 5492, 1295
\bibitem[Wu et al.(2011)]{WuJ11} Wu, J., Brandt, W.~N., Hall, P.~B., et al.\ 2011, \apj, 736, 28
\bibitem[Wu et al.(2012)]{WuJ12} Wu, J., Brandt, W.~N., Anderson, S.~F., et al.\ 2012, \apj, 747, 10
\bibitem[Xu et al.(2020)]{Xu20} Xu, X., Zakamska, N.~L., Arav, N., et al.\ 2020, \mnras, 495, 305
\bibitem[Yi et al.(2019a)]{Yi19a}Yi, W.; Brandt, W. N.; Hall, P. B., et al. 2019a, \apjs, 242, 28
\bibitem[Yi et al.(2019b)]{Yi19b}Yi, W.; Vivek, M.; Brandt, W. N., et al. 2019b, \apjl, 870, 25
\bibitem[Yi et al.(2020)]{Yi20} Yi, W., Zuo, W., Yang, J., et al.\ 2020, \apj, 893, 95
\bibitem[Yi \& Timlin(2021)]{Yi21} Yi, W. \& Timlin, J.\ 2021, \apjs, 255, 12
\bibitem[York et al.(2000)]{York2000} York, D.~G., Adelman, J., Anderson, J.~E., et al.\ 2000, \aj, 120, 1579
\bibitem[Zakamska \& Greene(2014)]{Zakamska14} Zakamska, N.~L. \& Greene, J.~E.\ 2014, \mnras, 442, 784
\bibitem[Zhang et al.(2014)]{Zhang14} Zhang, S., Wang, H., Wang, T., et al.\ 2014, \apj, 786, 42
\bibitem[Zuo et al.(2020)]{Zuo20} Zuo, W., Wu, X.-B., Fan, X., et al.\ 2020, \apj, 896, 


\end{thebibliography}
\end{document}